\documentclass[prl,twocolumn,superscriptaddress]{revtex4-2}

\usepackage{mathrsfs,amsmath,graphicx,paralist}
\usepackage[usenames]{color}

\usepackage[unicode=true,bookmarks=true,bookmarksnumbered=false,bookmarksopen=false,breaklinks=false,pdfborder={0 0 1},backref=false,colorlinks=true]{hyperref}

\hypersetup{linkcolor=magenta,urlcolor=blue,citecolor=blue,pdfstartview={FitH},hyperfootnotes=false,unicode=true}

    \def\<{\langle}         \def\>{\rangle}

  \def\V0{{\mathbf 0}}

  \def\B0{{\mathbf 0}}

\def\be{\begin{equation}}       \def\ee{\end{equation}}
\def\bea{\begin{eqnarray}}      \def\eea{\end{eqnarray}}

\begin{document}

\title{
Weyl Semimetal Made Ideal with  a Crystal of Raman Light and Atoms
}

\author{Xiaopeng Li}
\affiliation{Department of Physics, Fudan University, Shanghai 200433, China}

\author{W. Vincent Liu}
\email{wvliu@pitt.edu}
\affiliation{Department of Physics and Astronomy, University of
  Pittsburgh, Pittsburgh, Pennsylvania 15260, USA}
\affiliation{Wilczek Quantum Center, School of Physics and Astronomy and T. D. Lee Institute, Shanghai Jiao Tong University, Shanghai 200240, China}
        \affiliation{Shanghai Research Center for Quantum Sciences, Shanghai 201315, China}
        \affiliation{Shenzhen Institute for Quantum Science and Engineering,
          Southern University of Science and Technology, Shenzhen 518055, China}


\begin{abstract}

Optical lattices are known for their flexibility to emulate condensed matter physics and beyond. Based on an early theoretical proposal [{\it Science Bulletin} 65, 2080 (2020)], a recent experiment published { by Wang {\it et al.} 
[{\it Science} 372, 271 (2021)] }
accomplishes the first experimental realization  of topological band structure of the ideal Weyl semimetal in ultracold atomic matter,  prompting  fundamental interest in the context of gapless topological physics. With a neat design of 3D spin-orbit  interaction, the experiment has probed the gapless band topology through  spin texture imaging and quantum quench dynamics. This work has far reaching implications  to  topological effects and quantum anomaly in condensed matter and high energy physics.
\end{abstract}

\maketitle

\begin{figure*}[htp]
\includegraphics[width=.9\linewidth]{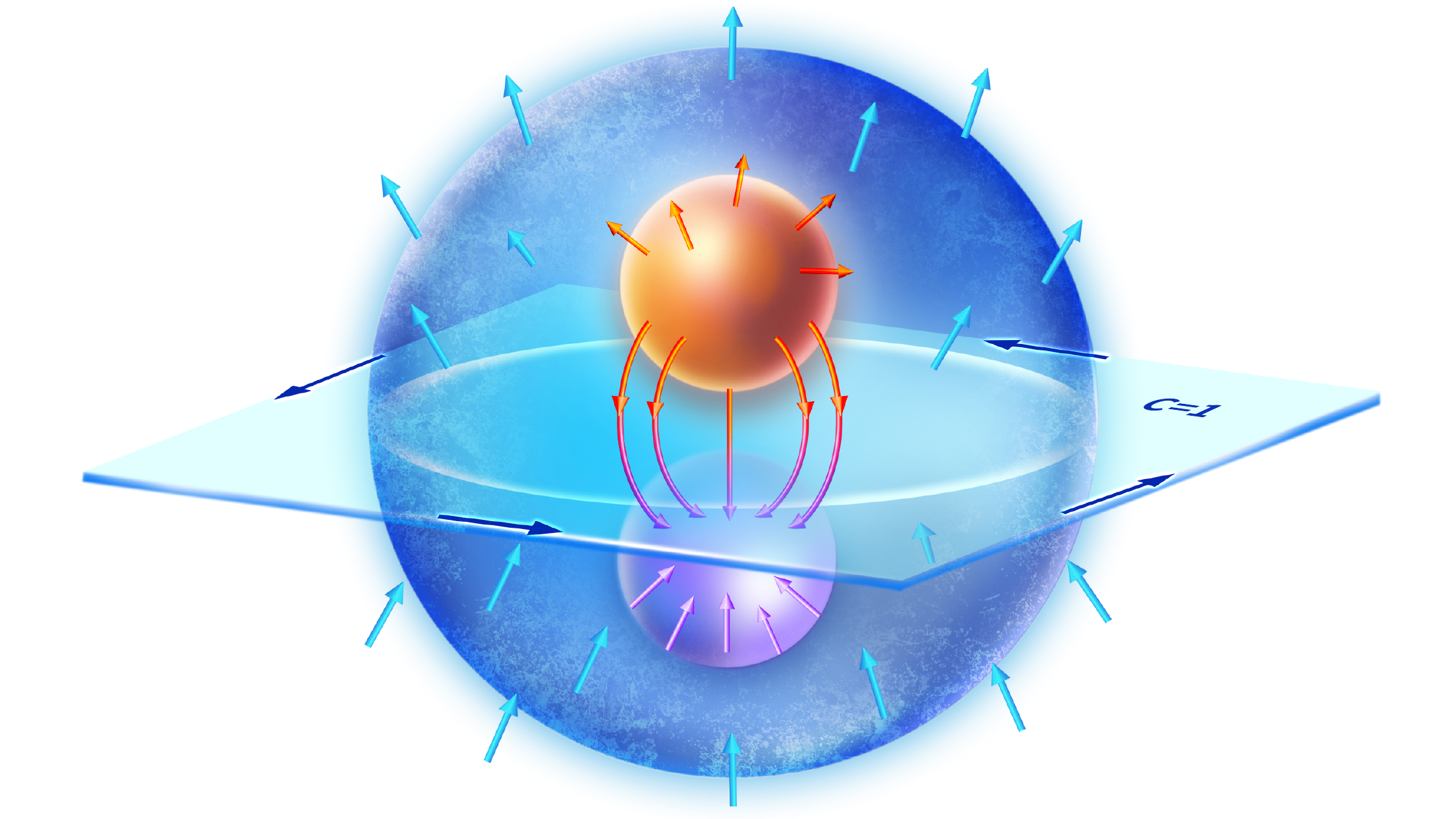}
\caption{Schematic illustration of ideal Weyl semimetal topology. 
Arrows in the plot illustrate the directions of Berry flux $\vec{\Omega}$, an analogue of magnetic fields in the momentum space. 
The two inner spheres enclose  two Weyl nodes with opposite chirality, representing the source and drain of Berry flux. The configuration of $\vec{\Omega}$ around the outer sphere  enclosing both Weyl nodes is topologically trivial. The horizontal plane separating the two Weyl nodes has a finite Chern number, giving rise to surface Fermi arc states. 
}
\label{fig:fig1}
\end{figure*}

Chiral symmetry played a spectacular role in the qualitative understanding of the low energy effective theory of quantum chromodynamics (QCD) and in the formulation of the standard model of electroweak and strong interactions in high energy physics. For the former, it is through the chiral symmetry breaking phenomena such as the emergence  of pions. For the latter, it provides constraints by requiring all axial-vector currents---generated by chiral symmetry transformation---be anomaly free and conserved when  coupled to any gauge field (see, for instance, ~\cite{Weinberg:bk05}).     This is the symmetry that left- and right-handed particles enjoy two independent sets of transformation while leaving all dynamics and physical observables invariant. Such particles were first introduced by Hermann Weyl in 1929 as a chirality-definite solution to Dirac equation at the massless limit.  In the family of elementary particles,  neutrinos which seemed to have vanishing mass in the beta decay experiment were initially expected to be Weyl fermions. This was later invalidated with mounting experimental evidence of non-zero neutrino masses. In condensed matter physics, Weyl fermions have been found to emerge as quasi-particle excitations of an exotic quantum many-body state of Weyl semimetals~\cite{2011_Wan_PRB}, with the vanishing effective mass protected by topological Berry phase effects in momentum space (Fig.~\ref{fig:fig1}), which has attracted tremendous research efforts in the last decade. Various exotic properties as mediated by Weyl fermions have been predicted including fascinating Fermi-arc surface states~\cite{2011_Wan_PRB}, chiral anomalous quantum transport~\cite{2015_Burkov_Science}, and topological phase transitions~\cite{2015_Bernevig_Nature}. In the first proposal of material realization with strongly correlated iridium pyrochlores R$_2$Ir$_2$O$_7$, 
twenty-four Weyl nodes were predicted with first principle calculations~\cite{2011_Wan_PRB}.  A conceptually much simpler and cleaner  proposal was provided based on stacking topological insulator thin films separated by ordinary-insulator spacer layers, leading to a most fundamental Weyl semimetal, 
{ which is an ideal Weyl semimetal having only two Weyl nodes,} 
the smallest possible number allowed in lattice models as proved by topological homotopy theory~\cite{1981_Nielsen_NPB}. 

For its simplicity 
{ the ideal Weyl semimetal with only two Weyl nodes has} 
been widely studied in theory. It has been established that the two nodes in the ideal Weyl semimetal cannot be trivially gapped out, being topologically protected. With interaction effects, an ideal Weyl semimetal either remains gapless or become gapped leaving exotic properties such as finite-momentum pairing~\cite{2012_Moore_PRB}, emergent space-time supersymmetry~\cite{2015_Yao_PRL}, or second Chern number protected chiral Majorana modes~\cite{2017_Liu_PRL}. In the last few years, there have been growing research efforts devoted to searching for ideal Weyl semimetals in magnetic topological materials~\cite{2019_Zhang_PRL}. However its realization with the approach of solid state material engineering has met various experimental obstacles  (e.g. sample quality upon magnetic doping),  with no direct observation made so far. 
 
\begin{figure*}[htp]
\includegraphics[width=\linewidth]{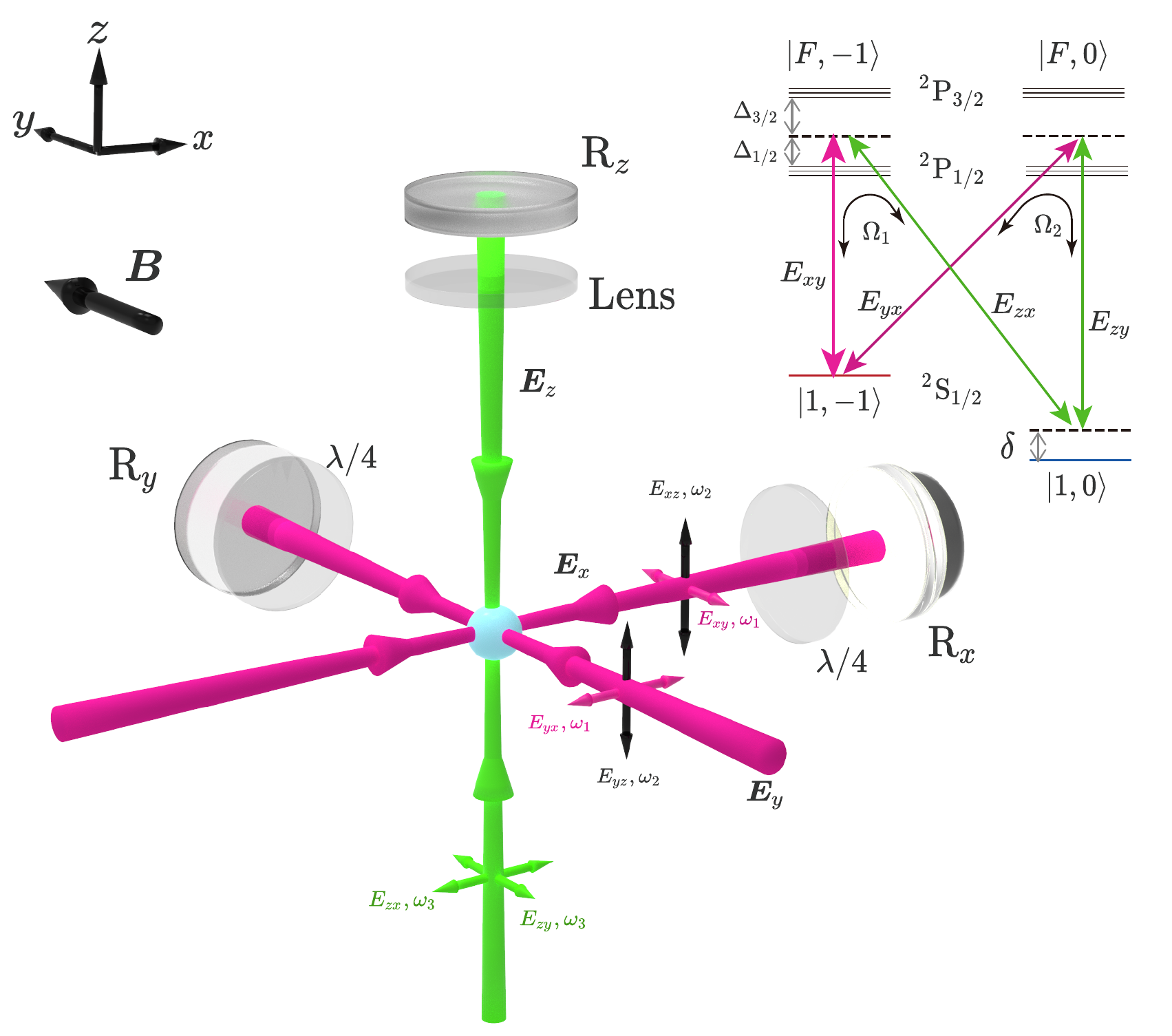}
\caption{  Experimental scheme of 3D Spin-Orbit coupling. Laser beams Ex, Ey, and Ez with a wavelength of $787$ nm produce a three dimensional light crystal with Raman couplings. The atoms make spin-flip hops between the nearest sites in the formed Raman light crystal. 
}
\label{fig:fig2}
\end{figure*}

In cold atom physics, there has been growing interest in achieving Wely semimetal through the route of time-reversal symmetry breaking. Rapid experimental developments have been made in creating synthetic spin-orbit interactions, providing a highly controllable platform to engineer such a chiral relativistic metal and other topological phases. The 1D spin-orbit interaction corresponding to Abelian gauge potential, when combined with conventional superconductivity, induces  effective $p$-wave pairing supporting Majorana zero modes, a potential building block for topological quantum computation. The 2D spin-orbit interaction corresponding to non-Abelian gauge potential facilitates even richer topological many-body states such as 2D topological insulators and chiral $p+ip$ superconductivity. Creating 3D spin-orbit interaction is one natural approach to engineer some Weyl semimetal Hamiltonian of chiral Lorentz-like energy spectrum. 
{ Synthesizing spin-orbit interactions from  one through two to three dimensions~\cite{2013_Galitski_Nature} represents} 
major steps towards quantum engineering of generic complex topological band structures, and at the same time the experimental challenges along the way drive cold atom-based quantum simulation technologies to flourish. After the 1D spin-orbit coupled Bose-Einstein condensate  was realized 
{in the cold atom experiment~\cite{2011_Spielman_Nature} with a Raman coupling scheme~\cite{2009_Liu_PRL},} 
it took about five years for synthesizing  the 2D spin-orbit interaction to succeed with an intriguing Raman optical lattice technique developed~\cite{2016_Pan_Science}.

After another five years, the experiment published in Science by Wang et al~\cite{2021_Pan_Liu_Science} has accomplished the realization of 3D spin-orbit interaction in cold atoms and reported the first evidence for the topological bands of ideal Weyl semimetal characteristics, 
based on the creative design put forward in early theoretical work by Lu et al~\cite{Lu-Wang-XJLiu:20}. The serial theoretical and experimental studies establish a remarkably  significant step furthering the quantum engineering of artificial topological phases. The experiment constructs a novel setup integrating a  Raman coupling-induced potential in a 3D optical lattice, 
by three retro-reflective laser beams with carefully designed laser frequencies (Fig.~\ref{fig:fig2}). The optical lattice is composed of a simple sinusoidal 1D lattice in one direction and a complex 2D checkerboard lattice in another two. 
The Raman potential created by two-photon processes, which bears a 2D character in continuous space, takes up a 3D form in the discrete coordinates of 
optical lattice~\cite{Lu-Wang-XJLiu:20}.
This neat design leads to  a stable 3D spin-orbit coupled Bose-Einstein condensate with Raman coupling-induced heating minimized.

In the experiment~\cite{2021_Pan_Liu_Science} the authors introduced two complementary approaches to successfully observe the ideal Weyl semimetal band structure. The first is a ``Virtual Slicing Image"  technique, mapping out the 3D spin texture through imaging a series of slices of 2D spin structures, by which two topological ``hedgehog"  defects with opposite chirality in the momentum space are clearly uncovered.  They further demonstrate the band topology through non-equilibrium quench dynamics---the second approach in their experiment, which double confirms the topological band structure supporting the claim of ideal Weyl semimetal.  

The experimental realization of 3D spin-orbit interaction has far reaching impact beyond the ideal Weyl semimetal in cold atom research frontiers. The achieved 3D spin-orbit coupled Bose-Einstein condensate provides a versatile platform to study the interplay of atomic interaction and non-Abelian gauge fields, which we expect to support intricate phases such as supersolids, unconventional spinor condensates, and complex magnetic textures, all being of fundamental interests in quantum many-body physics. It opens up wide opportunities to study three dimensional topological phenomena in general, especially in non-equilibrium settings. Merging the concept of topology with quantum dynamics has the potential of advancing the understanding of both subjects, towards new research disciplines.  We note that the experimental study ~\cite{2021_Pan_Liu_Science} is reported for a cold gas of $^{87}$Rb atoms, which are {\it bosonic}. It will be exciting to extend their idea of realizing 3D spin-orbit coupling and ideal Weyl semimetal to the case of fermionic atoms. 
That may create  a highly controllable quantum system to investigate chiral anomaly and chiral magnetic transport, which are fascinating topics being actively pursued by researchers in condensed matter physics. Further incorporating many-body interaction potentially produces artificial neutral pions as Goldstone bosons due to possible dynamical 
{ chiral symmetry breaking~\cite{Weinberg:bk05}.}
The experiment~\cite{2021_Pan_Liu_Science} should be appreciated as a timely milestone accomplishing Bose-Einstein condensation with ideal Weyl semimetal energy dispersion, whose rich physical properties  would at the same time inspire a wide range of future interesting directions at the multidisciplinary interface of cold atom, condensed matter and high energy physics.

{\bf Conflict of interests.}
The authors declare that they have no conflict of interest.

{\bf Acknowledgment} 
We acknowledge support by National Natural Science Foundation of China Grant No. 11774067 (X.L.),  by
AFOSR Grant No. FA9550-16-1-0006 and MURI-ARO Grant  No. W911NF17-1-0323 through UC Santa Barbara (W.V.L.), and by Shanghai Municipal Science and Technology Major Project Grant No. 2019SHZDZX01 (X.L.,W.V.L.).

\bibliography{Weylbib}

\end{document}